\documentclass[runningheads]{llncs}
\usepackage[T1]{fontenc}
\usepackage{dirtree}

\usepackage{cite}
\usepackage{amsmath,amssymb,amsfonts}
\usepackage{float}
\usepackage{hyperref}
\usepackage{algorithmic}
\usepackage{textcomp}
\usepackage{xcolor}
\usepackage{verbatim}
\usepackage{listings}
\usepackage{xcolor,colortbl}
\usepackage{todonotes}
\usepackage{tcolorbox}
\usepackage{csquotes}
\usepackage{ifthen}
\usepackage{caption}
\usepackage{subcaption}
\usepackage[ruled, lined, titlenotnumbered]{algorithm2e}

\usepackage{graphicx}
\usepackage{blindtext}

\usepackage{diagbox}
\usepackage{booktabs}
\usepackage{multirow}

\usepackage{pdflscape}
\usepackage{afterpage}
\usepackage{capt-of}

\usepackage{rotating}
\usepackage{xcolor,soul}
\usepackage{listings}
\usepackage{fontawesome}

\newcommand{\TwoSymbolsAndText}[3]{%
  #1\quad #2 \quad #3%
}

\sethlcolor{lightgray}
\tcbuselibrary{listings,skins,listingsutf8}
\definecolor{White}{rgb}{1,1,1}
\definecolor{codegreen}{rgb}{0,0.6,0}
\definecolor{codegray}{rgb}{0.5,0.5,0.5}
\definecolor{codepurple}{rgb}{0.58,0,0.82}
\definecolor{backcolour}{rgb}{0.95,0.95,0.92}
\definecolor{pblue}{rgb}{0.13,0.13,1}
\definecolor{lightblue}{rgb}{0.13,0.13,0.6}
\definecolor{pgreen}{rgb}{0,0.5,0}
\definecolor{pred}{rgb}{0.9,0,0}
\definecolor{pgrey}{rgb}{0.46,0.45,0.48}
\definecolor{mediumslateblue}{rgb}{0.48, 0.41, 0.93}
\definecolor{electricviolet}{rgb}{0.56, 0.0, 1.0}
\definecolor{LightCyan}{rgb}{0.88,1,1}
\definecolor{DarkBlue}{rgb}{0,0,0.2}
\definecolor{lightgreen}{rgb}{0.5,1,0.5}

\lstdefinestyle{mystyle}{
  backgroundcolor=\color{backcolour},   
  commentstyle=\color{codegreen},
  keywordstyle=\color{magenta},
  numberstyle=\tiny\color{black},
  stringstyle=\color{codepurple},
  basicstyle=\ttfamily\footnotesize,
  breakatwhitespace=false,     
  breaklines=true,         
  captionpos=b,          
  keepspaces=true,         
  numbers=left,          
  numbersep=5pt,          
  showspaces=false,        
  showstringspaces=false,
  showtabs=false,          
  tabsize=2
}
\usepackage[normalem]{ulem}
\useunder{\uline}{\ul}{}

\usepackage{amssymb}
\usepackage{pifont}

\newcommand{\cmark}{{\color{red}\ding{51}}}%
\newcommand{\xmark}{{\color{pgreen}\ding{55}}}%

\lstdefinestyle{myCustomIvyStyle}{
  aboveskip=0em,
  belowskip=0em,
  numbers=left,
  stepnumber=1,
  numbersep=20pt,
  tabsize=2,
  mathescape,
  basicstyle=\scriptsize,
  showspaces=false,
  showstringspaces=false,
  keywordstyle=\color{pblue},
  keywordstyle=[2]{\color{mediumslateblue}},
  keywordstyle=[3]{\color{electricviolet}},
  identifierstyle=\color{black},
  commentstyle=\itshape\color{pgreen},
  stringstyle=\color{pred},
  morekeywords={include,then,after,type,String,action,object,variant,of,import, instance, around, if, while, implement, else, before,ensure, module, returns, return,python, implementation},
  morekeywords=[2]{require,var,call, destructor,instantiate,struct, , execute,is_set,value,set,ip,ipv6,  interpret, virtual,relation, seconds, microseconds, milliseconds,function},
  morekeywords=[3]{this, stream_pos,stream_data,trans_params_struct,cid,quic_packet_type,bool,frame,stream_id,quic_packet,endpoint, microsecs, pkt_num},
  extendedchars=true,
  morecomment=[l]{\#}
}

\lstdefinestyle{myCustomIvyStyleHardcoded}{
  aboveskip=0em,
  belowskip=0em,
  firstnumber=10,
  numbers=left,
  roundcorner=10pt,
  stepnumber=1,
  numbersep=20pt,
  tabsize=2,
  basicstyle=\scriptsize,
  showspaces=false,
  showstringspaces=false,
  keywordstyle=\color{pblue},
  keywordstyle=[2]{\color{mediumslateblue}},
  keywordstyle=[3]{\color{electricviolet}},
  identifierstyle=\color{black},
  commentstyle=\itshape\color{pgreen},
  stringstyle=\color{pred},
  morekeywords={include,then,after,type,String,action,object,variant,of,import, instance, around, if, while, implement, else, before,ensure, module, returns, return,python, implementation},
  morekeywords=[2]{require,var,call, destructor,instantiate,struct, , execute,is_set,value,set,ip,ipv6,  interpret, virtual,relation, seconds, microseconds, milliseconds},
  morekeywords=[3]{this, stream_pos,stream_data,trans_params_struct,cid,quic_packet_type,bool,frame,stream_id,quic_packet,endpoint, microsecs, pkt_num},
  extendedchars=true,
  morecomment=[l]{\#}
}

\newtcblisting[use counter=inputPrg]{codeInput}[3]{
  listing only,
  listing style=myCustomIvyStyle,
  size=title,
  arc=1mm,
  enhanced jigsaw,
  coltitle=White,
  boxrule=0.5mm,
  title=\TwoSymbolsAndText{\faCode}{%
  \ifthenelse{\equal{#2}{}}{
  }{
  \textbf{\thetcbcounter. #2}
  }%
  }{\faCode},
  label=inputPrg:#3
}

\newtcblisting[use counter=inputPrg]{codeInputHard}[3]{
  listing only,
  listing style=myCustomIvyStyleHardcoded,
  size=title,
  arc=1mm,
  enhanced jigsaw,
  coltitle=White,
  boxrule=0.5mm,
  label=inputPrg:#3
}

\definecolor{Highlight}{HTML}{FFF59C}

\usepackage{tikz}

\begin{document}
\title{Formally Discovering and Reproducing Network Protocols Vulnerabilities}
\titlerunning{\textit{NACT}}
%
\author{Christophe Crochet \texttt{0000-0001-8635-2098}  \and John Aoga \texttt{0000-0002-7213-146X} \and
Axel Legay \texttt{0000-0003-2287-8925} }
\authorrunning{C. Crochet et al.}
\institute{
INGI, ICTEAM, Université catholique de Louvain, Place Sainte Barbe 2, L05.02.01, 1348 Louvain-La-Neuve, Belgium\\
\email{\{firstname.lastname\}@uclouvain.be}}
\maketitle              

\begin{abstract}
The rapid evolution of cyber threats has increased the need for robust methods to discover vulnerabilities in increasingly complex and diverse network protocols. This paper introduces \textit{Network Attack-centric Compositional Testing} (\textit{NACT})\cite{panther}, a novel methodology designed to discover new vulnerabilities in network protocols and create scenarios to reproduce these vulnerabilities through attacker models. \textit{NACT} integrates composable attacker specifications, formal specification mutations, and randomized constraint-solving techniques to generate sophisticated attack scenarios and test cases. The methodology enables comprehensive testing of both single-protocol and multi-protocol interactions. Through case studies involving a custom minimalist protocol (\texttt{MiniP}) and five widely used \texttt{QUIC} implementations, \textit{NACT} is shown to effectively identify, reproduce, and find new real-world vulnerabilities such as version negotiation abuse. Additionally, by comparing the current and older versions of these \texttt{QUIC} implementations, \textit{NACT} demonstrates its ability to detect both persistent vulnerabilities and regressions. Finally, by supporting cross-protocol testing within a black-box testing framework, \textit{NACT} provides a versatile approach to improve the security of network protocols. 

    \keywords{Formal Specifications, Formal Verification, Mutation Testing, Network Attacks, Systems, Internet protocols, \texttt{\texttt{QUIC}}, Concrete Implementation, Adverse Stimuli, Framework}
    
\end{abstract}

\section{Introduction}

In today’s hyperconnected world, network protocols serve as the foundation of global communication infrastructures, yet they remain vulnerable to sophisticated cyber threats. Protocol vulnerabilities provide a fertile ground for attackers to disrupt services, steal sensitive information, and cause widespread damage, as seen in high-profile incidents such as the \textit{Heartbleed}\cite{heartbleed14} and \textit{Log4Shell}\cite{everson2022log4shell} attacks. The unpredictability of cyberattacks lies in their ability to exploit weaknesses in ways not anticipated during the protocol's design. Attackers frequently leverage \textit{zero-day} vulnerabilities and subtle deviations from formal specifications to create unexpected behaviors that lead to breaches\cite{zeroday22}. This challenge highlights the necessity of anticipating attacks by uncovering potential vulnerabilities before they can be exploited~\cite{wen17}. Testing and simulation approaches play a crucial role in this proactive defense strategy.

Previous studies have delved into various fault-based testing methodologies. Some methods focus on validating protocols according to their defined specifications (conformance testing)~\cite{conformance2,conformance4}, while others target known vulnerabilities directly~\cite{vul5,vul10}, especially through mutation testing~\cite{9120695}, which creates test scenarios by injecting minor controlled errors (mutants) into protocols to gauge their resilience~\cite{mut7}. These strategies typically fall under the umbrella of \textit{black-box} or \textit{fuzz testing}~\cite{vul10}. It is crucial to perform tests directly on implementations to enhance the realism and applicability of these tests for real-world network situations. This is where \textit{Model-Based Testing (MBT)}~\cite{offutt1999generating} comes in, using an advanced language to describe protocol specifications and validate them directly against real implementations. This method facilitates the identification of potential code vulnerabilities that an attacker can exploit in practical scenarios, especially given the growing complexity of modern protocol implementations.

\sloppy{Recently, \textit{Network-centric Compositional Testing} (\textit{NCT}) \cite{McMillan_Zuck_2019} has emerged} as a pioneering MBT approach \cite{offutt1999generating} that uses formal protocol specifications, implemented with tools such as \textit{Ivy}~\cite{Padon_McMillan_Panda_Sagiv_Shoham_2016}, to generate test tools automatically. \textit{NCT} has effectively identified bugs and vulnerabilities, particularly in real-world protocols such as \texttt{QUIC}.

However, \textit{NCT} focused on individual compliance checks and did not address the dynamic and diverse nature of actual network interactions, which are essential to detect potential vulnerabilities before they can be exploited.

To bridge this gap, we present Network Attack-centric Compositional Testing (\textit{NACT}), a novel \textit{model-based mutation testing} that builds upon \textit{NCT}. \textit{NACT} is tailored to thoroughly evaluate real-world network protocols for security flaws. Using compositional attacker models, mutations in formal specifications, and randomized constraint-solving techniques, \textit{NACT} can uncover previously unknown vulnerabilities by emulating various advanced attack scenarios. Our method facilitates the examination of multi-protocol and cross-protocol interactions in black-box settings, thereby creating realistic, adversarial circumstances reflective of potential exploits in live environments. The capability to conduct multi-protocol testing and multipoint communications is becoming increasingly crucial with the rise of microservice architectures and distributed systems. We validate our method's effectiveness by testing several recent vulnerabilities in the widely used \texttt{QUIC} protocol, successfully identifying new significant security weaknesses, including version negotiation issues and buffer overflows.

The remainder of the paper is organized as follows. Section~\ref{sec:back} offers an overview of network protocol vulnerabilities, associated attacks, the \textit{NCT} methodology, and its inherent limitations. Section~\ref{sec:approach} introduces the \textit{NACT} methodology, elaborating on the mutations of formal specification, their implementation, and integration within a compositional testing framework. Section~\ref{sec:exp} discusses our case studies on five \texttt{QUIC} implementations, showcasing the ability of \textit{NACT} to identify complex vulnerabilities. Finally, Section~\ref{sec:conclusion} concludes and suggests possible directions for future work.

\section{Background}\label{sec:back}

This section covers the core concepts of network attacks, how to test network protocols against these threats using network-centric approaches, and the limitations of these approaches, emphasizing the need for more comprehensive testing frameworks.

\paragraph{\textbf{(1) Network Attack Vectors and Agents.}}
In network security, a \textbf{vulnerability} is a defect or weakness present in the design, implementation, or operation of a system that an attacker can exploit to breach the system's security attributes, such as \textit{confidentiality}, \textit{integrity}, \textit{availability}, and \textit{authenticity}. When a malicious entity discovers and exploits a vulnerability, it results in a \textbf{network attack}, which is any intentional action to disrupt standard system functioning.

\paragraph{(a) Types of Network Attacks.} To understand the different forms of network attacks, we conducted a literature review and identified the most prevalent types of attacks that can occur in network environments. We classify these attacks according to their mode of operation. Table \ref{tab:attacks} summarizes the general categories of network attacks as referenced in various works \cite{iot_formal,HOQUE2014307,8462745,quic_attack}.

\begin{table}[h!]
\centering
\begin{tabular}{|c|l|}
\hline
\textbf{Type of Attack} & \multicolumn{1}{c|}{\textbf{Description}}                                                                                          \\ \hline
\begin{tabular}[c]{@{}c@{}}TLS/SSL\\Connection\\Attacks\end{tabular}   &
  \begin{tabular}[c]{@{}l@{}}Exploiting vulnerabilities in the TLS handshake process, such as \\ protocol downgrade or renegotiation attacks, to intercept, decrypt, \\ or alter encrypted communication data.\end{tabular} \\ \hline
Injection               & \begin{tabular}[c]{@{}l@{}}Injecting malicious data/code into a system or application to \\ manipulate its operations.\end{tabular} \\ \hline
Eavesdropping           & \begin{tabular}[c]{@{}l@{}}Intercepting communications between two parties without their\\ consent.\end{tabular}                  \\ \hline
\begin{tabular}[c]{@{}c@{}}Service \\ Disruption\end{tabular} &
  \begin{tabular}[c]{@{}l@{}}Overloading or disrupting services to make them unavailable to \\  legitimate users.\end{tabular} \\ \hline
Spoofing                & Impersonating a legitimate entity to deceive users or systems.                                                                     \\ \hline
\begin{tabular}[c]{@{}c@{}}Protocol-Based \\Attacks\end{tabular} &
  \begin{tabular}[c]{@{}l@{}}Exploiting inherent weaknesses or misconfigurations in com-\\ munication protocols (e.g., TCP Slowloris) to overwhelm, dis-\\ rupt services, or gain unauthorized access.\end{tabular} \\ \hline
\end{tabular}
\caption{Network attacks categories from the literature.}
\label{tab:attacks}
\end{table}

\noindent These attack types form the basis for our testing framework, allowing us to simulate various attack scenarios and validate the robustness of network protocols under different adversarial conditions.

\paragraph{(b) Attack Vectors.} An \textbf{attack vector} is the specific method or pathway that an attacker uses to exploit a vulnerability in a system. Understanding attack vectors is essential in constructing effective security models and testing methodologies, as they directly influence how attacks are simulated and mitigated. We identified three main attack vectors relevant to network security testing, presented in Table~\ref{tab:attack_vectors}. 
Although malicious clients and servers might appear similar, the directions of their interactions differ. A malicious client begins communication and strays from protocols to perform attacks. In contrast, a malicious server responds to requests with harmful actions, taking advantage of client trust. The impact of a malicious client is precise and targeted, whereas a malicious server can influence numerous users and disrupt the entire network, magnifying the damage. 
A special case of this is the Man-in-the-Middle (MitM) attack, where the attacker intercepts and manipulates the communication between a client and server, often impersonating both simultaneously. This dual role allows the attacker to perform activities such as spoofing, eavesdropping, and data injection without the communicating parties' awareness.

\begin{table}[h!]
\begin{tabular}{|c|c|c|c|}
\hline
\textbf{Attack Vector} &
  \textbf{Description} &
  \textbf{Attack Surface} &
  \textbf{Type of Attack} \\ \hline
\begin{tabular}[c]{@{}c@{}}Man-in-\\ the-Middle \\ (MitM)\end{tabular} &
  \begin{tabular}[c]{@{}c@{}}Intercepts and \\ potentially alters the \\ communication between \\ two parties without their \\ knowledge, compromising \\ message integrity \\ and authenticity.\end{tabular} &
  \begin{tabular}[c]{@{}c@{}}Network \\ communication \\ between client \\ and server\end{tabular} &
  \multirow{3}{*}{\begin{tabular}[c]{@{}c@{}}Eavesdropping\\ Injection\\ TLS/SSL \\ Connection\\ Protocol-Based \\ Attacks\\ Service Disruption\\ Spoofing\end{tabular}} \\ \cline{1-3}
\begin{tabular}[c]{@{}c@{}}Malicious \\ Client\end{tabular} &
  \begin{tabular}[c]{@{}c@{}}Behaves as \\ a legitimate user but \\ intentionally violates \\ protocol rules to disrupt \\ the system or steal data.\end{tabular} &
  \begin{tabular}[c]{@{}c@{}}Interaction with \\ a legitimate \\ server\end{tabular} &
   \\ \cline{1-3}
\begin{tabular}[c]{@{}c@{}}Malicious \\ Server\end{tabular} &
  \begin{tabular}[c]{@{}c@{}}Controls a seemingly legitimate\\ server that responds \\ to client requests \\ maliciously, violating \\ protocol rules and \\ compromising security.\end{tabular} &
  \begin{tabular}[c]{@{}c@{}}Responses to \\ legitimate client \\ requests\end{tabular} &
   \\ \hline
\end{tabular}
\caption{Identified Attack Vectors in Network Protocols}
\label{tab:attack_vectors}
\end{table}

\noindent These attack vectors serve as the foundation for developing our \textbf{Network Attack Model Framework}. In this framework, we emulate different attack scenarios to assess network protocols for potential vulnerabilities. Further details will be provided in the next section. Now, let's explain the Network-Centric Testing framework that underpins our methodology.

\paragraph{\textbf{(2) Network-Centric Testing Approaches.}}\label{subsec:nct}
Network-centric compositional testing (\textit{NCT}) is a specialized approach within model-based testing (\textit{MBT}) that focuses specifically on network protocols. The principle of compositional testing views formal specifications as a system of interrelated components or processes, each with its own input and output. This method allows for examination of the protocol behaviors in their network interactions, instead of relying exclusively on abstract mathematical models. The emphasis on genuine network behavior is what distinguishes the methodology as "network-centric".

\paragraph{{\textit{NCT} Principle.}} \textit{NCT} creates a systematic framework for generating formal specifications of Internet protocols and verifying their implementations for conformity to these specifications~\cite{McMillan_Zuck_2019,Crochet_Rousseaux_Piraux_Sambon_Legay_2021}. After generating the formal model code, a test generator produces concrete and randomized test cases. The testers, which use an \textit{SMT solver} to meet the constraints set by the formal protocol requirements, are then deployed to verify the real-world implementation of the protocol. If any protocol requirements are not met, the resulting traces can be examined to find and identify potential errors or vulnerabilities. Let us consider a minimal network protocol designated as \texttt{MiniP} to illustrate how \textit{NCT} works.

\paragraph{{MiniP.}} The Minimalist Protocol (\texttt{MiniP}) is specified as follows. Each packet must contain exactly two frames. There are three types of frames : \texttt{PING}, \texttt{PONG}, and \texttt{TIMESTAMP} frames. The \texttt{PING} and \texttt{PONG} frames include a \textit{four-byte string} that corresponds to the word 'ping' or 'pong', respectively. A packet must contain one of these two frames. Additionally, the \texttt{TIMESTAMP} frame contains an \textit{eight-byte unsigned integer} that represents the time, in milliseconds, when the packet is transmitted. Every packet includes this frame. The client initiates communication by sending a packet with the \texttt{PING} frame and the \texttt{TIMESTAMP} frame as its payload. The server is required to reply within \textit{three seconds} with a packet that includes the \texttt{PONG} frame followed by the \texttt{TIMESTAMP} frame. This process repeats until the client disconnects. To end the connection, the client simply stops sending packets for \textit{more than three seconds}.

\begin{figure}[h!]
  \centering
  \includegraphics[width=0.97\columnwidth]{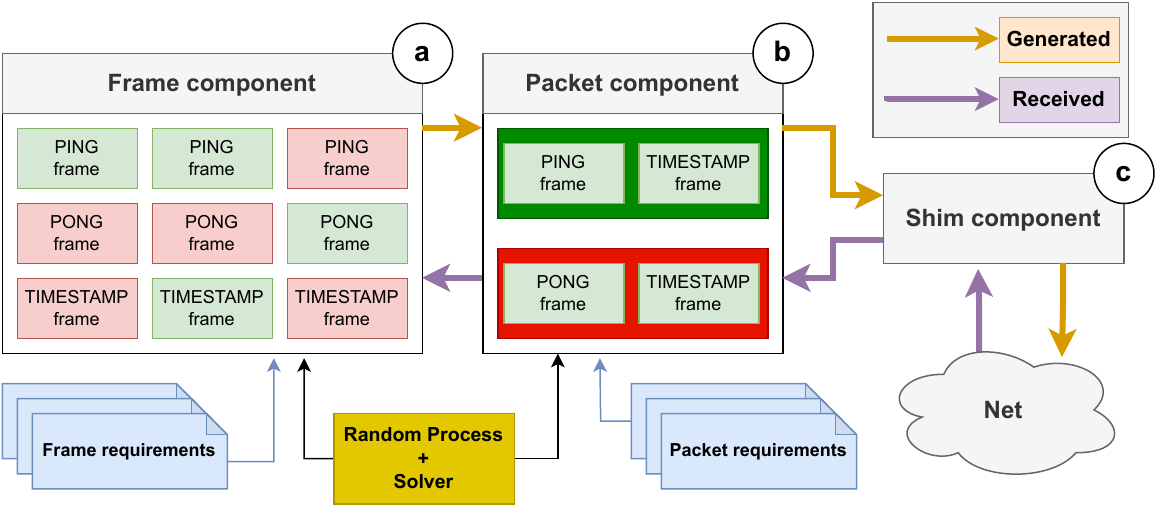}
  \caption{\texttt{MiniP} Network-Centric Testing (\textit{NCT}) structure}
  \label{fig:ncm}
\end{figure}

 \paragraph{{MiniP Network-centric Structure.}} Figure~\ref{fig:ncm} illustrate the design of \texttt{MiniP} according to the \textit{NCT} principle. The "Frame" process generates output that is used as input for the "Packet" process. The assumptions regarding the inputs of a process are treated as guarantees for the outputs of other processes. Each element represents a layer of the \texttt{MiniP} stack, including the frame layer \textcircled{a} and the packet layer \textcircled{b}. The \textit{shim} component \textcircled{c} handles the network transmission and packet reception. Once a packet is received, the shim component checks the formal specifications associated with the packet. For example, it ensures that a MiniP packet consistently includes two frames in the appropriate sequence. Should any of the criteria not be satisfied, an error is triggered. Frames are likewise handled according to their own formal specifications.

\paragraph{{Formal specification.}} \textit{NCT} employs the Ivy language~\cite{Padon_McMillan_Panda_Sagiv_Shoham_2016,McMillan_Padon_2020} for formal specification purposes. Ivy facilitates describing a program's state through first-order logic formulas and incorporates relations (boolean predicates), functions, modules, and type objects as the primary abstractions to model the system's state. For example, the Ivy code snippet~\ref{inputPrg:ping} defines how the handling of the \texttt{PING} frame, which must contain the word 'ping' as a \textit{four-byte string}.

\begin{codeInput}{c}{Ping frame formal specification code}{ping} 
before ping_frame.ping.handle {
    require f.data     = ping_data; #data must be 'ping'
    require f.data.end = 4; #length of the data must be 4
}
\end{codeInput}

\noindent While \textit{NCT} approaches have been effective in identifying protocol compliance issues, it is crucial to extend them to include tests that explore beyond safety properties focus, as outlined in their paper~\cite{McMillan_Zuck_2019a}.

\paragraph{\textbf{(3) Network-Centric Testing Limitations.}} \textit{NCT} and Specification-conforming testing, which focuses on ensuring that implementations strictly adhere to formal specifications, has several inherent limitations:

\paragraph{{\normalfont\textcircled{1}} Limited to Specification-Conforming Testing.} Focusing solely on testing that adheres to the specifications restricts the overall extent of coverage. By strictly following the formal specifications, the methodology might overlook potential issues stemming from unexpected or non-specification-conforming inputs and behaviors, which are prevalent in practical scenarios. This aspect is crucial, as real-world networks frequently encounter unforeseen interactions or attacks that are not predicted by a specification. This limitation underscores the necessity to broaden the testing framework to include tests that go beyond the specified boundaries, possibly incorporating formal attack models that mimic adversarial conditions or unusual network behaviors.

\paragraph{{\normalfont\textcircled{2}} Single Protocol Testing.} The current approach is designed to test one protocol at a time. This approach limits the framework’s applicability to more complex systems where multiple protocols interact, such as in distributed services or multi-layered web applications. These interactions might introduce unforeseen issues or vulnerabilities. Expanding the methodology to support multi-protocol testing would significantly improve its ability to detect interaction-based vulnerabilities. 

\paragraph{{\normalfont\textcircled{3}} Restricted to Point-To-Point communications.}  The testing methodology is currently limited to point-to-point communications, which does not fully capture the complexity of modern networked systems that often involve multipoint or broadcast communications. Many real-world network applications require the verification of protocols that manage communications among multiple nodes, such as in mesh networks or distributed systems. Extending the testing framework to handle these more complex communication patterns would significantly enhance its usefulness and applicability.

\paragraph{\textbf{Putting all together.}}  Addressing these limitations by integrating \textit{formal non-conform (mutation) attack models}, supporting \textit{multi-protocol environments}, and expanding to\textit{ multi-node communications} would greatly enhance the robustness and comprehensiveness of protocol testing, ensuring that the protocol performs well not only in controlled environments but also in real-world, dynamic network scenarios.

\section{Network Attack Model Framework}\label{sec:approach}

\textit{Network Attack-centric Compositional Testing (\textit{NACT})}, a method of model-based mutation testing, facilitates the evaluation of network protocols' security and robustness by introducing mutations (minor alterations) to the specifications (\textit{addressing Limitation }\textcircled{1}) and testing these modifications across different use cases (\textit{addressing Limitations }\textcircled{2} \& \textcircled{3}). To accomplish this, we suggest a framework for a network attacker model that operates in three phases.

\paragraph{\textbf{Step1: Defining formal specification mutations.}}
Mutation testing has been successfully applied to network protocols to improve their robustness and security. This approach involves intentionally making small changes (mutations) to the protocol specifications, resulting in mutant versions that might reveal potential weaknesses\cite{BUDD198563,167603,5209814,12901}.
Subsequently, the mutants are evaluated against a comprehensive set of test cases to determine the efficacy of existing tests in detecting artificially introduced errors. The main goal is to evaluate the protocol's fault tolerance and to determine potential failure points that may arise in unexpected circumstances. 

Table~\ref{tab:mutation_techniques} presents various mutation techniques applicable to formal specifications in internet protocol testing. These specific mutations are chosen for their ability to effectively challenge and validate the robustness of protocol implementations. 

Furthermore, they specifically target critical protocol behavior aspects that are prone to reveal hidden issues. Other mutations might not be effective in this scenario, either due to insufficiently significant deviations or the generation of invalid mutants that fail to yield meaningful test results.

\begin{table}[h!]
\centering
\begin{tabular}{|p{0.15\textwidth}|p{0.28\textwidth}|p{0.30\textwidth}|p{0.26\textwidth}|}
\hline
    \centering\textbf{Mutation} & 
    \centering\textbf{Description} & 
    \centering\textbf{Problem Investigated} & 
    \textbf{Challenges} \\ \hline
    \centering\textbf{\centering\textbf{Statement Deletion or Addition}} & 
    Remove or add statements to observe protocol deviations & 
    Functional errors, missing logic, incomplete flows, Protocol deviations, unintended side-effects & 
    Incomplete protocol states, trivial bugs, unreachable or invalid states \\ \hline 
    
     \centering\textbf{Negation Mutation} & 
    Negate logical expressions to introduce faults & 
    Logical errors, incorrect conditional handling & 
    Reveal deep logical errors \\ \hline    

    \centering\textbf{Value Replacement} & 
    Modify variables or constants to test behavior under changes & 
    Edge-cases, faulty variable handling & 
    Maintain meaningful tests \\ \hline

    \centering\textbf{Boundary Value Mutation} & 
    Alter values to boundary limits for edge-case testing & 
    Buffer overflows, boundary issues & 
    Select boundaries ca- refully to avoid meaningless mutations \\ \hline
    \centering\textbf{Control Flow Mutation} & 
    Modify control structures to test robustness & 
    Conditional logic failures, Infinite loops, performance bottlenecks & 
    Unreachable states, complex conditions, computational cost\\ \hline  
    \centering\textbf{Mutate Data Structures} & 
    Change data structure definitions to find vulnerabilities & 
    Memory management, invalid state transitions & 
    Ensure valid structure mutations \\ 
\hline
\end{tabular}
\caption{Formal Specification Mutations in Internet Protocol Testing.\label{tab:mutation_techniques}}
\end{table}

\begin{figure}[h!]
  \centering
  \includegraphics[width=1\columnwidth]{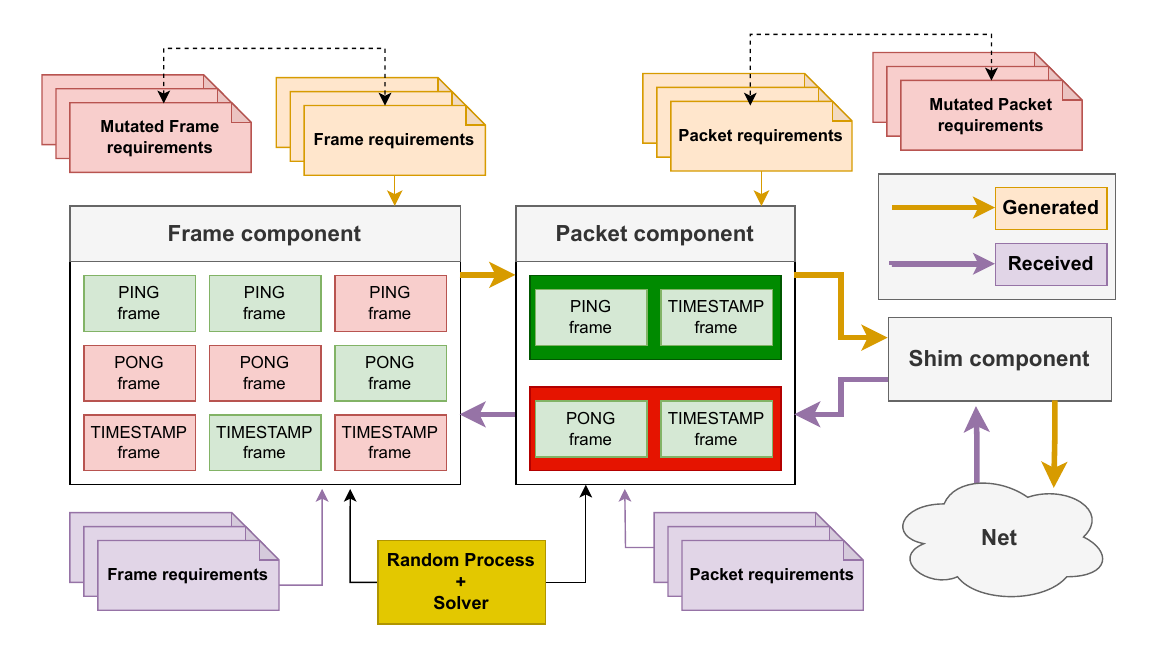}
  \caption{\texttt{MiniP} Formal Specification Mutation}
  \label{fig:mutated-model}
\end{figure}

At the protocol layer, these mutations are incorporated into both the frame and packet components. Figure~\ref{fig:mutated-model} illustrates the integration of mutations within \texttt{MiniP}'s architecture. Based on the requirements, modifications are made to define the mutations, which are subsequently embedded in the protocol components, thereby exploring how the implementation handles these deviations from the standard specification (\textit{Limitation} \textcircled{1}).

\noindent To test these mutations effectively, we must now implement them. This is where \textit{NCT} becomes relevant.

\paragraph{\textbf{Step2: Implementing composable attacker specifications.}}
Our approach for implementing mutations extends the conventional formal specification mutation testing framework by introducing mutations directly into the formal specification. This is facilitated by \textit{NCT}. Currently, generating mutants is a manual process due to the difficulty of generalizing mutations across all network protocols while maintaining consistent and meaningful system behavior. To ensure a degree of generality, we provide a template for each attack model: the man-in-the-middle, malicious client, and server (Table~\ref{tab:attack_vectors}). 

At a protocol level, we show below, using our running protocol \texttt{MiniP}, how to implement a malicious character injection (a mutation by addition), find possible security vulnerabilities (vs safety vulnerabilities) in this mutant, and implement a test scenario to test the vulnerability.

\paragraph{\textbf{(1) Finding vulnerability with specification mutation.}} 
The specific mutation introduced here consists in altering a requirement to allow the inclusion of potentially harmful characters in the packet payload. The following Ivy code snippet \ref{inputPrg:pingmalicious} represents this mutation showing the modification of the original ping frame handler presented in the code snippet \ref{inputPrg:ping} Sect.\ref{subsec:nct}.

\begin{codeInput}{c}{Format String \& Buffer Overflow}{pingmalicious} 
before ping_frame.ping.handle_maliciously {
    require f.data.end < 100000 & exists I. I < f.data.end-8  & 
      (f.data.value(I)   = 0x25  & f.data.value(I+1) = 0x78  &
       f.data.value(I+2) = 0x25  & f.data.value(I+3) = 0x6e);
}
\end{codeInput}

\noindent This mutation is designed to evaluate the \texttt{MiniP}'s robustness against format string vulnerabilities, a well-known category of security issues. In addition, we varied the size of the message payload to assess the protocol’s handling of unusually large or small messages, a common technique for identifying buffer overflow vulnerabilities. This vulnerability is particularly severe, as it can result in memory corruption, leading to potential system crashes or arbitrary code execution by an attacker. Once a vulnerability is identified using randomized mutation testing, the next step involves creating specific scenario test cases.

\paragraph{\textbf{(2) Exploiting vulnerability creating a test scenario}}
 A scenario test case is designed to explicitly target the identified vulnerabilities and structured to replicate the exact conditions under which the vulnerability was found. 
 The test scenarios are created using attack vectors (Table~\ref{tab:attack_vectors}). For example, we can replicate the protocol's behavior during a \textit{Man-in-the-Middle attack} or its reaction to \textit{malicious clients} sending improperly formatted packets under mutations. By incorporating these vectors into mutation testing, we can validate the system's ability to withstand real-world attacks. These scenario-based tests can be reused in future tests to ensure that the vulnerability remains mitigated in subsequent implementations or protocol versions. One can use \textit{NCT} to craft test scenarios with straightforward assertion directives. The detected vulnerabilities (string format and buffer overflow) can be tested using the following code snippet (\ref{inputPrg:pingscenario}).
 \begin{codeInput}{c}{Malicious Character Injection Specific test}{pingscenario} 
before ping_frame.ping.handle_maliciously {
    require f.data.end = 50; #Buffer overflow size + format string
    require f.data.value(6) = 0x25 & f.data.value(7) = 0x78; 
}
\end{codeInput}

\noindent Now that we define mutations and test scenarios, we need to test the protocol under test in an appropriate environment. 

\paragraph{\textbf{Step3: Simulating attack scenarios within virtual networks.}}
A crucial stage in developing and validating network protocols involves thorough testing to confirm that protocols operate as expected under various network conditions. Our methodology highlights the significance of testing in settings that closely resemble real-world scenarios. Consequently, we mainly use \textit{virtual networks}. 

\begin{figure}[h!]
  \centering
  \includegraphics[width=1\columnwidth]{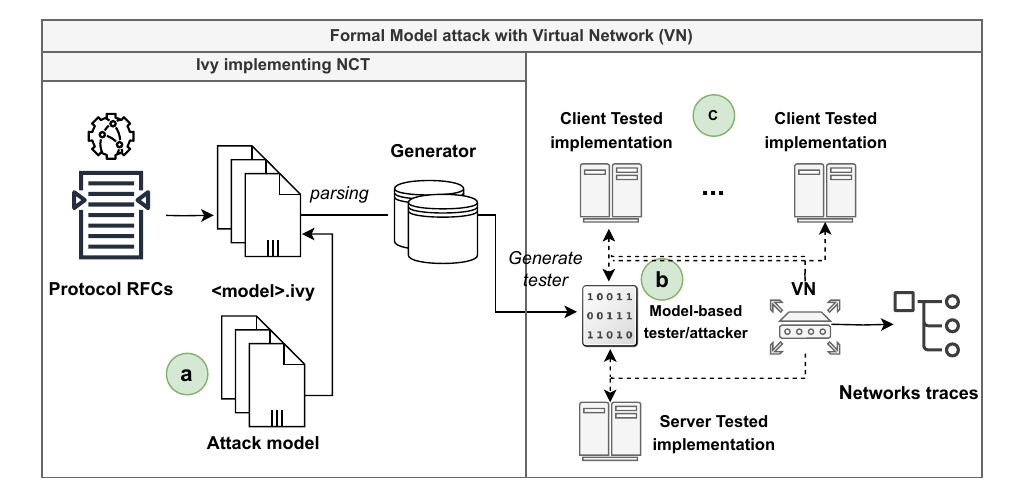}
  \vspace*{-0.5cm}
  \caption{Virtual networks and \textit{NACT}}
  \label{fig:fsm2}
\end{figure}

\paragraph{\textbf{(1) Using virtual networks.}} Virtual networks, by leveraging namespaces, enable the creation of isolated network environments that accurately emulate actual deployment settings. This approach facilitates the testing of multipoint communications (\textit{Limitation} \textcircled{3}) and complex topologies, imitating real-world network dynamics. Building upon \textit{NCT}, our framework is compatible with simulators. These simulators provide various benefits, including reproducibility, accurate management of network conditions, and the ability to simulate intricate scenarios such as timing attacks~\cite{nsct24}. However, employing simulators necessitates the creation of simulated system calls (syscalls) to emulate the network stack, which can be challenging in proportion to the complexity of the simulation condition. Extending \textit{NCT} to complex network systems, especially in microservices architectures, involves dealing with the added complexity of interacting protocols.
Figure~\ref{fig:fsm2} illustrates the interaction between our framework and a virtual network. The \textit{NCT} framework integrated with an attack model \textcircled{a} produces model-based testers or attackers \textcircled{b}, which are used within the virtual network consisting of multiple client-tested implementations \textcircled{c} interfaced with a server-tested implementation. Upon completion of the simulation, the network traces are analyzed to investigate vulnerabilities.

\begin{figure}[h!]
  \centering
  \includegraphics[width=0.9\columnwidth]{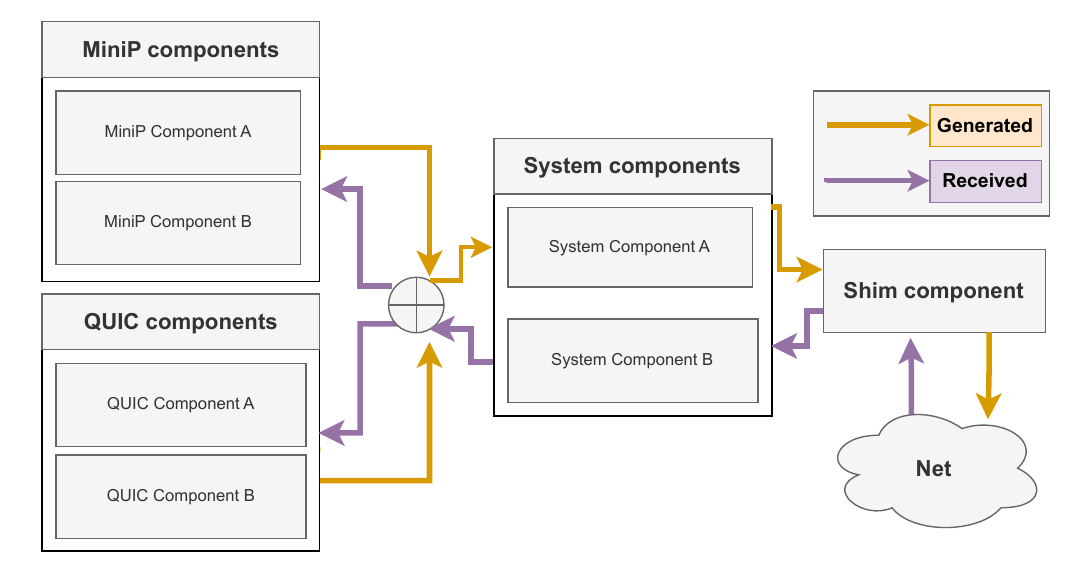}
  \vspace*{-1em}
  \caption{System Component}
  \label{fig:mutated-system}
\end{figure}

\paragraph{\textbf{(2) Compositional Testing of Networks Services.}} \textit{NCT} originally focused on testing single protocols through formal specifications and automated testers, ensuring compliance via \textit{SMT solvers} \cite{McMillan_Zuck_2019}. In a microservice environment, where services may use different protocols (e.g., HTTP over TCP, gRPC over \texttt{QUIC}), the challenge is to develop modular formal specifications that capture both individual service behavior and inter-service interactions \cite{netsketch_tool}. To do so, we design the framework with some level of abstraction, illustrated in Figure~\ref{fig:mutated-system}. The \textit{system component}, a virtual component, serves as a master framework that defines base events common across all protocols, ensuring consistent integration and testing (\textit{Limitation} \textcircled{2}). The model can then be expanded with \textit{protocol-specific requirements}, allowing each protocol to be tested both in isolation and as part of the broader system. A \textit{shim} layer manages interactions between different protocols, routing events to the appropriate components and ensuring that protocol-specific requirements are applied while maintaining system integrity.  

We implemented a toy example demonstrating a \textit{MitM} attack scenario where \texttt{QUIC} packets are reflected towards \texttt{MiniP} endpoints ending in buffer overflow. This simple setup provides a valuable scenario into the interaction between different protocols under attack conditions, highlighting potential vulnerabilities in cross-protocol communication environments. 

\vspace*{-2em}

\vspace*{-2.5em}

\paragraph{\textbf{Putting all together.}} The proposed Network Attack-centric Compositional Testing (\textit{NACT}) framework provides a structured and comprehensive approach to enhancing the security and robustness of network protocols. 
This framework integrates the mutation of formal specifications, the deployment of attacker specifications that can be composed, and the simulation of attack scenarios in virtual network environments to provide a flexible approach for uncovering hidden vulnerabilities. Our method performs specification-conforming \textit{and} non-conforming testing, managing both individual protocol and multiple protocol interactions in multipoint communication settings. 

\vspace{1em}
In the subsequent section, we will demonstrate how this framework is applied to the testing of the real world \texttt{QUIC} protocol and its efficacy in uncovering new vulnerabilities.

\afterpage{%
    \clearpage
    \thispagestyle{empty}
    \begin{landscape}
        \centering 
        \resizebox{0.96\linewidth}{!}{
\begin{tabular}{|p{0.04\linewidth}|p{0.16\linewidth}|p{0.2\linewidth}|p{0.2\linewidth}|p{0.15\linewidth}|p{0.25\linewidth}|}
    \hline
    \centering\textbf{N°} &
    \centering\textbf{Vulnerability} & 
    \centering\textbf{Description} & 
    \centering\textbf{Known Causes} & 
    \centering\textbf{Occurred Problems} & 
    \centering\textbf{Attack Scenario} \tabularnewline \hline
    
    \centering\textbf{(1)} &
    CVE-2022-30591  \texttt{QUIC} Slow Loris \cite{a2022_nvd}  & 
    DoS via incomplete \texttt{QUIC} or HTTP/3 requests  & 
    Misparsing of the MTU Discovery service by the \textit{\texttt{quic-go}} implementation  & 
    CPU consumption and DoS via \textit{Slowloris} attack   & 
    Maintain the connection with minimal data transfer, forcing the server to hold resources indefinitely \cite{10.1007/978-981-99-1312-1_7} \tabularnewline \hline

    \centering\textbf{(2)} &
    CVE-2023-42805  \texttt{QUIC} Frame Parsing Exploitation \cite{a2023_nvd}  & 
    Incorrect parsing of unknown \texttt{QUIC} frames causing erroneous responses  & 
    Reception of unknown frames within \texttt{QUIC} packets prior to fix   & 
    Erroneous server responses   & 
    Exploit frame parsing vulnerabilities to manipulate the server response \\ \hline
    
    \centering\textbf{(3)} &
    CVE-2024-22189  \texttt{QUIC} Memory Exhaustion \cite{a2024_nvd}  & 
    Memory exhaustion through excessive transmission of \texttt{NEW\_CONNECTION\_ID} frames   & 
    Exploitation of congestion control by collapsing the window   & 
    Denial of service, memory exhaustion   & 
    Flood the server with requests to retire connection IDs, overwhelming the server’s resources \\ \hline

    \centering\textbf{(4)} &
    GitHub issues  \textit{\texttt{picoquic}} Malicious Payload Injection \cite{privateoctopus_2020_denial}  & 
    Malicious \texttt{QUIC} frame triggers an infinite loop, causing a DoS  & 
    Crafting maliciously \texttt{QUIC} frames processed during session epoch 3   & 
    Infinite loop, server crash, remote exploitation   & 
    Inject crafted payloads to trigger an infinite processing loop \\ \hline
    
    \centering\textbf{(5)} &
    GitHub issues  \texttt{quant} Malicious Payload Injection \cite{ntap_2020_dos}  & 
    DoS caused by processing connections with identical connection IDs  & 
    Duplicating connection IDs processed by \texttt{quant}   & 
    Server crash and denial of service   & 
    Inject frames to cause server failure and potentially execute malicious payloads \\ \hline

    \centering\textbf{(6)} &
    Non-conformance bug in \texttt{QUIC} Implementations (\texttt{NEW\_TOKEN} frame) \cite{10.1145/3664476.3664521}  & 
    Several \texttt{QUIC} implementations (\texttt{picoquic} 7f2fbdfb \& \texttt{lsquic} 3.2.0) fail to handle invalid packets properly  & 
    Protocol specification non-conformance   & 
    Invalid error codes and server instability   & 
    Replay malformed \texttt{NEW\_TOKEN} frames to reproduce server errors and non-conformance \\ 
\hline
\end{tabular}}
\captionof{table}{\texttt{QUIC} selected vulnerabilities and our defined attack scenarios}
\label{tab:attack-scenarios}
    \end{landscape}
    \clearpage
}

\section{Experiments}\label{sec:exp}

This section is organized into two parts. The first part focuses on reproducing known vulnerabilities in various \texttt{QUIC} \cite{9000} implementations, such as \texttt{picoquic}, \texttt{quic-go}, \texttt{quinn}, \texttt{quant}. We apply established attack scenarios to both current and previous versions of these implementations to verify the effectiveness of security patches and identify any regressions. The second part involves the discovery of new vulnerabilities using our Network Attack-centric Compositional Testing (NACT) framework on \texttt{lsquic} and \texttt{quant} implementation.

\paragraph{\textbf{(1) Testing recent \texttt{QUIC} vulnerabilities}.} To evaluate vulnerabilities within our framework, it is required to define both the vulnerabilities and the associated attack scenario, given the mutations are defined (Section~\ref{sec:approach}). Table~\ref{tab:attack-scenarios} shows the chosen \texttt{QUIC} vulnerabilities and the respective attack scenario designed to address them. 

\begin{table}[h!]
\centering
\begin{tabular}{|c|c|c|c|}
\hline
\textbf{Implementation}                        & \textbf{Language} & \textbf{SLOC} & \textbf{Version}  \\ \hline
\cite{picoquic} \texttt{picoquic} (a)          & C                 & 122k          & bb6799 (actual)   \\ \hline
\texttt{picoquic} (b)                          & C                 & 84k           & 42c620 (draft 27) \\ \hline
\cite{quicgo_2024_github} \texttt{quic-go} (a) & Go                & 83k           & v0.46.0           \\ \hline
\texttt{quic-go} (b)                           & Go                & 73k           & b5ef99a           \\ \hline
\cite{quinnrs_2024_github} \texttt{quinn} (a)  & Rust              & 33k           & 0.11.2            \\ \hline
\texttt{quinn} (b)                             & Rust              & 41k           & 0.10.0 (draft 29) \\ \hline
\cite{ntap_2016_github} \texttt{quant} (a)     & C                 & 18k           & dc7721            \\ \hline
\texttt{quant} (b)                             & C                 & 18k           & bf903d (draft 29) \\ \hline
\end{tabular}
\caption{Tested implementations informations}
\label{tab:implems}
\end{table}

To thoroughly assess the security of \texttt{QUIC} implementations in relation to these vulnerabilities, we performed tests on both the latest and older versions of several widely used libraries. Each \texttt{QUIC} implementation was examined for known vulnerabilities using its most recent version as well as a previous version to detect any regressions or persistent issues. 
By testing multiple versions, we ensure that vulnerabilities identified in earlier versions are indeed present and can be reliably reproduced, while also verifying that these vulnerabilities have been effectively fixed in the most recent versions, preventing their reintroduction.

To guarantee the dependability and consistency of our results, each test was executed three times across the different \texttt{QUIC} implementations. Due to the variability inherent in some tests, this approach allowed us to control for any inconsistencies in the data and better evaluate the robustness of each implementation against the identified vulnerabilities. 
Table~\ref{tab:results} displays the summary of these tests' outcomes. For each test, we allocated a time budget of 150 seconds per test \cite{panther_ivy_result}.

\begin{table}[h!]
\centering
\resizebox{\columnwidth}{!}{%
\begin{tabular}{|c|cc|cc|cc|cc|c|c|}
\hline
\textbf{\diagbox{Attacks}{Implem.}} &
  \textbf{\begin{tabular}[c]{@{}c@{}}\texttt{picoquic}\\ (a)\end{tabular}} &
  \textbf{\begin{tabular}[c]{@{}c@{}}\texttt{picoquic}\\ (b)\end{tabular}} &
  \textbf{\begin{tabular}[c]{@{}c@{}}\texttt{quic-go}\\ (a)\end{tabular}} &
  \textbf{\begin{tabular}[c]{@{}c@{}}\texttt{quic-go}\\ (b)\end{tabular}} &
  \textbf{\begin{tabular}[c]{@{}c@{}}\texttt{quinn}\\ (a)\end{tabular}} &
  \textbf{\begin{tabular}[c]{@{}c@{}}\texttt{quinn}\\ (b)\end{tabular}} &
  \textbf{\begin{tabular}[c]{@{}c@{}}\texttt{quant}\\ (a)\end{tabular}} &
  \textbf{\begin{tabular}[c]{@{}c@{}}\texttt{quant}\\ (b)\end{tabular}} \\ \hline
  \textbf{(1)} &
  \xmark &
  \xmark &

  \xmark &
  \cellcolor[HTML]{808080} $\sim$ &
  \xmark &
  \xmark &
  \xmark &
  \xmark \\ \hline
\textbf{(2)} &
  \xmark &
  \xmark &

  \xmark &
  \xmark &
  \xmark &
  \cellcolor[HTML]{808080} $\sim$ &
  \cmark &
  \cmark \\ \hline
\textbf{(3)} &
  \xmark &
  \xmark &

  \xmark &
  \cellcolor[HTML]{808080} $\sim$ &
  \xmark &
  \xmark &
  \xmark &
  \cmark \\ \hline
\textbf{(4)} &
  \xmark &
  \cellcolor[HTML]{808080}\cmark &

  \xmark &
  \xmark &
  \xmark &
  \cmark &
  \xmark &
  \cmark \\ \hline
\textbf{(5)} &
  \xmark &
  \xmark &

  \xmark &
  \xmark &
  \xmark &
  \xmark &
  \xmark &
  \cellcolor[HTML]{808080}\cmark \\ \hline
\textbf{(6)} &
  \xmark &
  \cellcolor[HTML]{808080}\cmark &
  
  \xmark &
  \xmark &
  \xmark &
  \cmark &
  \xmark &
  \cmark \\ \hline \hline
\textbf{\textit{Results}}  & \textit{safe} & \textit{unsafe} & \textit{safe} & \textit{unsafe} & \textit{safe} & \textit{unsafe} & \textit{unsafe} & \textit{unsafe} \\\hline
\end{tabular} 
}
\caption{Reproduction of Vulnerability Testing Results for Various QUIC Implementations. QUIC Implementations versions: (a) \textit{actual} and (b) \textit{old}. Each symbol represents the outcome of the tests, running three times: \cmark{} indicates that the three trials failed (vulnerable), \xmark{} indicates that the three trials passed (safe), and $\sim$ indicates mixed results (some trials passed and some failed). The attacks numbers (i), 1 to 6, are described in Table~\ref{tab:attack-scenarios}. \hl{Grey cells} represent the implementation(s) originally detected as vulnerable to the attack scenario. An implementation is considered \textit{safe} if it passes all tests and \textit{unsafe} otherwise.}
\label{tab:results}
\end{table}

The experiments carried out provided valuable information on the behavior and vulnerabilities of various \texttt{QUIC} implementations. As shown in Table~\ref{tab:results}, our testing revealed that while some implementations demonstrated resilience against specific attack scenarios, others exposed vulnerabilities that could lead to potential security risks. Notably, our findings highlighted the effectiveness of newer versions in addressing previously identified issues, showcasing the ongoing efforts to enhance the security and robustness of \texttt{QUIC} implementations.

The results indicate that while certain implementations, such as \texttt{quic-go} (a), showed no significant issues, others, like \texttt{quinn} (b) and \texttt{quant} (b), revealed persistent vulnerabilities or improper handling of edge cases, as detailed in the various test outcome.

One of the key observations was the effectiveness of newer versions in addressing previously identified vulnerabilities. For example, in the case of \texttt{picoquic}, the vulnerabilities observed in version (b), which included issues concerning the frame parsing process and RFC compliance, were successfully mitigated in the newer version (a), demonstrating a clear improvement in security.

However, the experiments also highlighted some persistent challenges. Certain implementations, such as \texttt{quinn} (b), showed issues in handling edge cases, such as repeated sending of \texttt{CONNECTION\_CLOSE} frames ($> 10$) followed by Stateless Reset packets, indicating a need for further refinement. Similarly, \texttt{quant} (b) frequently experienced crashes or improper connection closures when dealing with oversized tokens, illustrating the difficulties in achieving robust \texttt{QUIC} implementations across different scenarios.

The experiments faced some difficulties, particularly in consistently triggering specific vulnerabilities like the Slowloris attack on \texttt{quic-go} (b). This challenge underscores the complexity of certain attack vectors and the ambiguous nature of some CVE designations, which can make reproducibility and consistent testing outcomes difficult.

In conclusion, these results underscore the significance of comprehensive and rigorous testing frameworks for the assessment of protocol security. The varying degrees of resilience and vulnerability observed across different versions and implementations underscore the necessity for continuous vigilance, rigorous testing, and the refinement of security measures. By addressing these challenges, the security and robustness of \texttt{QUIC} and similar network protocols can be significantly enhanced, ensuring their resilience against evolving threats.

\paragraph{\textbf{(2) Discovering new network protocol vulnerabilities}.}
In the course of our investigation into \texttt{QUIC} protocol implementations, we identified a number of vulnerabilities that could potentially be exploited under certain conditions. This section presents our findings, with a particular focus on scenarios involving MitM and client-initiated attacks.

\paragraph{\textbf{(2.1) Version Negotiation Abuse - Denial of Service}}
The LiteSpeed \texttt{QUIC} (\texttt{lsquic}) Library is a fast, flexible, and production-ready open-source implementation of \texttt{QUIC} and HTTP/3 for servers and clients. It has been used in LiteSpeed products since 2017 and supports multiple \texttt{QUIC} versions. With around 129k SLOC, \texttt{lsquic} is significantly larger compared to other implementations. A vulnerability was identified with regard to the version negotiation process. In particular, when \texttt{lsquic} initiates a handshake using version 0xff000022 (draft-34) and receives a version negotiation packet responding with 0xff00001d (draft-29), the implementation erroneously applies an incorrect encryption key to subsequent packets. This results in checksum failures, which could potentially lead to a denial of service (\textit{DoS}) condition.

It is noteworthy that this issue does not manifest when transitioning between certain other versions, such as from 0xff00001b (draft-27) to 0xff00001d (draft-29). However, the vulnerability is consistently manifested when moving from version 0xff000022 (draft-34) to 0xff00001b (draft-27). Testing the reverse transition (from 0xff00001d to 0xff000022) is challenging with the current \texttt{lsquic} setup, as it defaults to selecting the highest available version.

The root cause of this vulnerability is not related to the key generation process, as each version employs a distinct salt for deriving encryption keys. Rather, the issue arises from the inappropriate application of these keys following version negotiation. This vulnerability presents a potential vector for a MitM attack. While initial packets are encrypted, they do not fully protect against MitM attacks. Furthermore, version negotiation packets, as defined in RFC9000, are entirely unprotected. Consequently, an attacker could intercept and manipulate the version negotiation packet, either by altering the selected version or crafting a fraudulent packet. This could potentially lead to a denial-of-service (DoS) attack.

This vulnerability appears to be specific to the \texttt{lsquic} implementation and is not necessarily indicative of a flaw in RFC9000 itself. However, the RFC's relatively vague guidelines on the version negotiation process could contribute to implementation inconsistencies, leading to potential vulnerabilities like the one identified, where an off-path attacker could exploit the protocol's flexibility to deplete server resources.

\paragraph{\textbf{(2.2) Malformed Frame Encodings - Memory exhaustion attacks Exploits}}

In the course of our preliminary experiments with a modified specification, we identified a significant error in the implementation of \texttt{QUIC}'s \texttt{quant} (a) functionality. In particular, when a specific type of frame encoding was employed, the \texttt{quant} process entered an infinite loop during the connection close procedure. This issue was particularly evident when the client transmitted a \texttt{NEW\_TOKEN} frame, which is not permitted by the RFC, in conjunction with a parsing error in the HTTP request contained within the \texttt{STREAM} frame. These errors resulted in an infinite loop, as the frame encodings differed slightly from the anticipated format but remained within the acceptable range, prompting the \texttt{quant} algorithm to classify them as valid input. Consequently, the connection close state was not correctly exited, leading the server to continue consuming resources and potentially resulting in a system crash.

A subsequent test with a different mutation of the specification yielded comparable results, though through a distinct mechanism. In this scenario, the altered frame encoding modified certain fields within the \texttt{NEW\_CONNECTION\_ID} frame header in a way not anticipated by the original specification. The parsing error in the HTTP request exacerbated the situation, leading to repeated, unsuccessful attempts by the "\texttt{quant}" process to close the connection. The unexpected packet structure caused the connection closure process to enter an indefinite loop.

In both cases, the infinite loop was triggered when the server initially sent a \texttt{CONNECTION\_CLOSE} frame with a \texttt{PROTOCOL\_VIOLATION} error code twice, followed by an endless sequence of HTTP 505 errors encapsulated within \texttt{FLOW\_CONTROL\_ERROR} frames. In the context of the mutated specification and introduced parsing errors, this combination of frames resulted in the persistent failure of the connection termination process, which ultimately caused the server to become unresponsive.

Notably, this issue does not appear in older versions of \texttt{quant} (b), suggesting that it has been introduced in more recent updates. This highlights the importance of continuous testing across different versions to identify and address newly introduced vulnerabilities.

\section{Conclusion \& Future works}\label{sec:conclusion}

Robust network protocol testing is crucial in an interconnected digital world due to evolving cyber threats. Our research explores attack vectors and develops a comprehensive, mutation-based testing framework to identify vulnerabilities. We addressed several attacks by constructing the \textit{Network Attack-centric Compositional Testing (\textit{NACT})} framework. Mutations generate adversarial conditions by modifying protocol specifications, exposing hidden vulnerabilities that traditional tests might miss, especially within formal specification testing (\textit{NCT}).

While \textit{NCT} has laid the groundwork for testing protocol compliance, it inherits some limitations. One inherited challenge is its focus on specification conformance, which leaves out adversarial behavior and multi-protocol interactions. Another challenge is the point-to-point communication limitation missing the complexity of modern distributed systems. We addressed these limitations by incorporating non-conformant mutation-based attack models, supporting multi-protocol environments, and extending the framework to multi-node communications using virtual networks. These enhancements enable the framework to perform more comprehensive security testing across diverse network environments.

Our experimental results with \texttt{QUIC} demonstrate the efficacy of our framework. We developed consistent attack scenarios to validate and discover vulnerabilities, notably in version negotiation and frame encoding errors, leading to \textit{denial of service (DoS)} or \textit{infinite loop conditions}. These findings underscore the need for robust security testing frameworks for both known and emerging vulnerabilities in protocols such as \texttt{QUIC}.
 
Looking ahead, our research aims to enhance the \textit{NACT} framework by incorporating the \textit{Network Simulator-centric Compositional Testing} (\textit{NSCT}) methodology~\cite{nsct24}. NSCT has built upon NCT to verify time-dependent network properties, ensuring reproducibility of experiments via simulators. This approach will aid in replicating attacks and allow the modeling of timing attacks. 

We also intend to conduct extensive empirical testing on real network systems associated with the AMC3 project.\cite{amc3}.

\paragraph{\textbf{Acknoledgements}} We would like to thank the belgium's "\textit{Defence-related Research Action}" (DEFRA) and the "\textit{Automated Methodology for Common Criteria Certification}" project (AMC3).

\newpage
\bibliographystyle{splncs04}
\bibliography{bib}
\end{document}